# Higher harmonic resistance oscillations in micro-bridge superconducting Nb ring

Masashi Tokuda[1]*, Ryoya Nakamura[1], Masaki Maeda[1], and Yasuhiro Niimi[1, 2]**

[1]Graduate School of Science, Department of Physics, Osaka University, Toyonaka, Osaka 560-0043, Japan
[2]Center for Spintronics Research Network, Osaka University, Toyonaka, Osaka 560-8531, Japan
*tokuda@meso.phys.sci.osaka-u.ac.jp, **niimi@phys.sci.osaka-u.ac.jp

We studied resistance oscillations in two types of superconducting mesoscopic Nb rings. In a simple superconducting ring device, a resistance oscillation with a period of the quantized magnetic flux $h/2e$ was clearly observed. On the other hand, in a micro-bridge ring device where two-narrow parts are embedded in parallel and work as superconductor-normal metal-superconductor junctions, higher harmonic resistance oscillations were obtained when the measurement current was well-tuned. We argue that such higher harmonic resistance oscillations can be detected even in the micro-bridge Nb superconducting ring device where the device size is much larger than the coherence length of Nb.

Mesoscopic superconducting devices provide an excellent platform for investigating phase-related phenomena, owing to the macroscopic wave functions[1]. In a Josephson junction where two superconductors are weakly connected via a very thin insulator, normal metal, or micro-bridge structure, a supercurrent flows depending on the relative phase difference between the two superconductors without any voltage applied[2-4]. When two Josephson junctions are connected in parallel in a superconducting ring, the magnetic flux through the ring is quantized by $\phi_0 = h/2e$, where $h$ is Planck's constant and $e$ is the electric charge. This is well-known as superconducting quantum interference device (SQUID) and has been utilized as a very sensitive magnetic field sensor[5].

When the distance $d$ between two weak links is comparable or shorter than the coherence length $\xi$ of superconductor, unique phase interference phenomena can be expected because of the interaction between the junctions via order parameter variations, quasiparticle diffusion[6] and Andreev bound state[7]. For superconducting Al ring-shaped devices with two normal metal junctions where $d$ is comparable to $\xi$ of Al, second harmonic resistance oscillations with a period of $h/4e$ have been reported[8]. However, the detailed mechanism of such intriguing resistance oscillations has not been elucidated yet. Even for devices with $d \gg \xi$, similar higher harmonic patterns can be seen[9,10], depending on the shunt resistance, capacitance, and loop inductance in the superconducting ring-shaped devices[3]. From the viewpoint of application, such higher harmonic components have to be suppressed in order to realize the normal operation of SQUID[11], but on the contrary, a high-performance magnetic sensor can be expected by controlling the higher harmonic components[8]. In this sense, further experimental and theoretical studies on harmonic oscillations in superconducting devices with $d \gg \xi$ and higher critical temperatures $T_c$ such as Nb are desirable.

In this work, we measured resistance oscillations in two types of superconducting Nb ring-shaped devices, i.e., simple ring device and micro-bridge ring device with two weak links. In both cases, a clear resistance oscillation with a period of $h/2e$ was observed. Only for the micro-bridge ring device, much more complicated resistance oscillations were detected. From fast Fourier transform (FFT) analysis, it turned out that up to the fourth harmonics were included in the resistance oscillations. Such higher harmonic resistance oscillations have never been reported in Nb superconducting ring-shaped devices where the perimeter is much larger than $\xi$ of Nb.

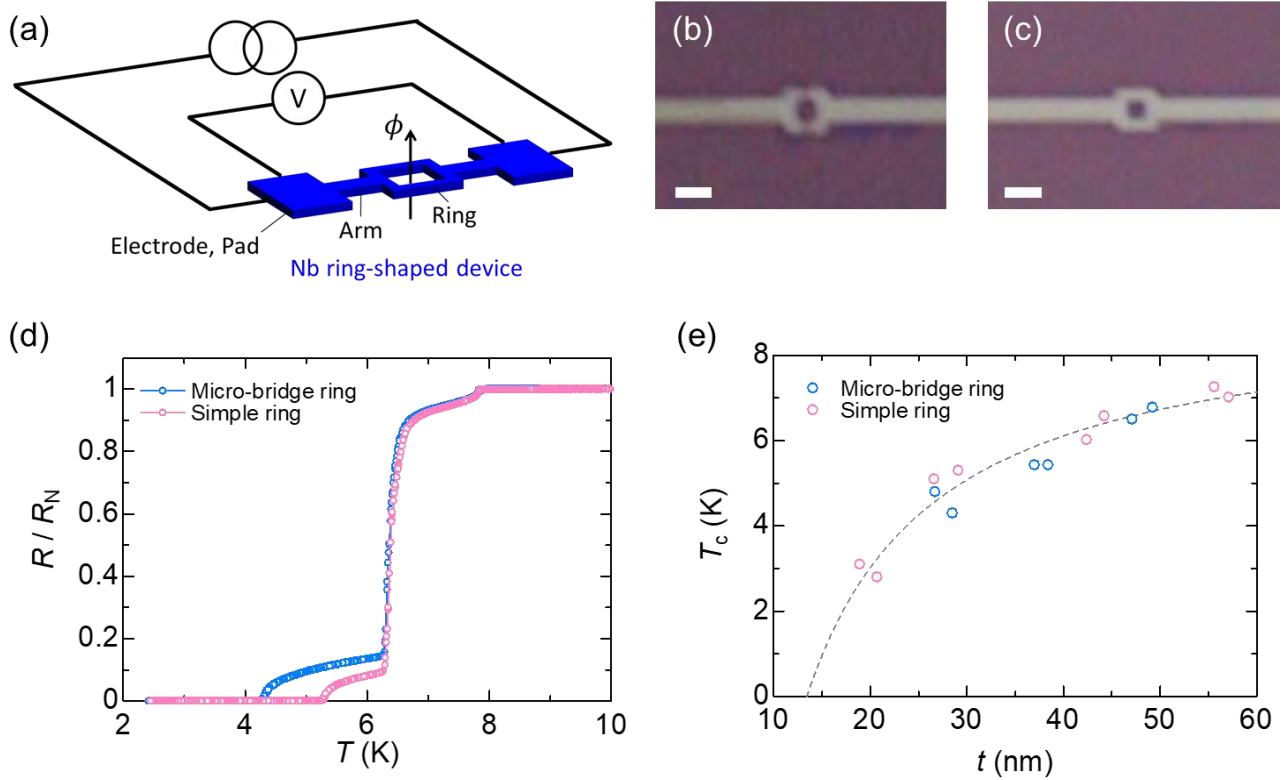

**Fig. 1. (a)** Schematic of the device structure and measurement circuit. **(b), (c)** Optical microscope images of micro-bridge ring device and simple ring device, respectively. The scale bars correspond to 1 µm. **(d)** Temperature dependence of resistance for the two types of ring-shaped devices. The vertical axis is normalized by the resistance in the normal state $R_N$ at 8 K. **(e)** Thickness dependence of $T_c$. The dashed line expresses a guide to the eye.

Two types of ring-shaped devices, i.e., simple ring and micro-bridge ring devices, were fabricated using Nb. We first patterned the two types of ring-shaped devices on a SiO$_2$/Si substrate coated with polymethyl methacrylate resist, using electron beam lithography. The desired patterns were obtained by depositing Nb thin films by means of dc magnetron sputtering and removing the rest of the patterns with the lift-off technique.

Standard four-terminal measurements were performed with an ac lock-in amplifier except for dc current-voltage ($I$-$V$) measurements. Figure 1(a) shows the schematic drawing of the device structure and the measurement circuit. The samples were cooled down from room temperature to about 2 K using a $^4$He flow refrigerator. The external magnetic field was applied with an electromagnet and measured with a Hall sensor.

In Figs. 1(b) and 1(c), we show optical microscope images of the micro-bridge ring and simple ring devices, respectively. In both cases, the thickness of the ring part is $t \approx 30$ nm, while the thickness of other parts (such as arms and electrodes in Fig. 1(a)) is $t \approx 40$ nm. These values were measured with an atomic force microscopy. Such a thickness difference can often be seen in devices fabricated by using magnetron sputtering and the lift-off technique. The micro-bridge ring device has two narrow parts in parallel, which behave as superconductor-normal metal-superconductor (SNS) junctions because the superconductivity of the narrow parts becomes weaker than the other parts. The width and the length of the narrowest part are 300 nm, while the width of the other parts and also the simple ring are kept at 600 nm.

Typical resistance versus temperature curves for the two types of devices are shown in Fig. 1(d). Let us first focus on the critical temperatures $T_c$ of the two devices. In the present work, $T_c$ is defined as the temperature when the resistance of the ring becomes zero. Although $T_c$ of bulk Nb is about 9 K, $T_c$ of our Nb thin films strongly depends on the thickness, as reported in Refs. 12)-15). Figure 1(e) shows the thickness $t$ dependence of $T_c$ of our ring-shaped devices. Here $t$ is defined as the thinnest thickness in the ring part. Although $T_c$ is as low as 3 K for $t = 20$ nm, it approaches $T_c$ of bulk Nb with increasing $t$. Such a thickness dependence of $T_c$ has been well-established[16,17] and reported not only in Nb but also in other superconducting thin films[18-21].

Now we come back to the temperature dependence of resistance for the two types of rings. The resistance is normalized by that in the normal state $R_N$ at $T = 8$ K. It starts to drop at about 7.8 K, exhibits a sharp drop at 6.3 K, and becomes zero at 5.3 (simple ring) and 4.3 K (micro-bridge ring). Such a multi-step transition in the $R$ vs $T$ curves would be due to the nonuniform superconducting transition temperature of the devices. As mentioned above, the thickness and width ($w$) of the device are slightly different, depending on the segments, i.e, electrode, arm, and ring. The first resistance drop at 7.8 K corresponds to the superconducting transition of electrodes and pads ($w > 10$ μm). With decreasing temperature, the resistance gradually decreases due to the proximity effect penetrating from the Nb electrodes to the arms ($w = 1$ μm). The arms become superconducting at 6.3 K, resulting in the sharp resistance drop. Thus, the residual resistance below this temperature corresponds to the resistance of the ring and the micro-bridge part. The residual resistance vanishes as the temperature decreases further, but $T_c$ is obviously different between the two cases. The difference stems from the narrow parts in the micro-bridge device.

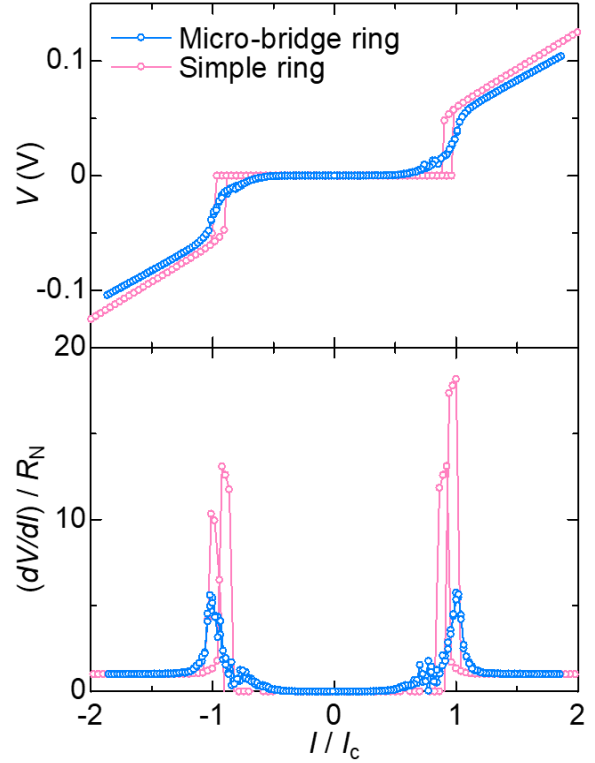

**Fig. 2.** Upper panel: Current-voltage ($I$-$V$) properties of simple ring and micro-bridge ring devices measured at 2.5 K. Lower panel: Differential resistances obtained from the $I$-$V$ curves. The vertical axis is normalized by the normal state resistance $R_N$. Both horizontal axes are normalized by the critical current $I_c$ defined as the current where the differential resistance takes the maximum.

In order to see the resistance difference between the simple ring and micro-bridge ring devices more clearly, we performed $I$-$V$ measurements at 2.5 K, as shown in the upper panel of Fig. 2. The horizontal axis is normalized by the critical current $I_c$ which is defined as the maximum point of the differential resistance obtained from the $I$-$V$ curve shown in the lower panel of Fig. 2. The critical currents of the simple ring and micro-bridge ring devices are 0.25 and 0.2 mA, respectively. The smaller critical current of the micro-bridge ring is reasonable because the micro-bridge ring has two weak links and the critical temperature is lower by 1 K, compared to that of the simple ring, as shown Fig. 1(d). The simple ring device shows a sharp transition with a hysteresis at $I_c$. The hysteretic behavior near $I_c$ originates from the heating effect. On the other hand, the transition observed in the micro-bridge ring device is broader. This broad transition in the micro-bridge device is due to the weak superconductivity at the narrow parts. Moreover, such an $I$-$V$ characteristic with no hysteresis is peculiar to a resistively shunted SQUID[3]. Therefore, the narrow parts in the micro-bridge ring device work as weak links, i.e., SNS junctions.

Next, we measured the magnetic field dependence of resistance with various measurement currents. Figures 3(a) and 3(b) show clear resistance oscillations observed in the micro-bridge ring and simple ring devices, respectively. The periods $B_0$ are in good agreement with those estimated from $B_0 = \phi_0/S$, where $S$ is the surface of the ring. $B_0 \approx 1$ mT for the micro-bridge ring device and $B_0 \approx 1.5$ mT for the simple ring device. When the measurement current $I$ is small enough compared to $I_c$, the resistance keeps zero because the superconductivity is robust. As $I$ approaches $I_c$, the superconductivity becomes weaker and a finite resistance appears. As a result, the resistance oscillates, depending on the external magnetic field. While in the simple ring device, a standard resistance oscillation was observed at $I \lesssim I_c$, more complicated resistance oscillations appeared in the micro-bridge ring device when $I$ was set to $0.93 \sim 0.94 I_c$. When $I \sim 0.97 I_c$, such complex resistance oscillations almost disappeared. We confirmed similar resistance oscillations for other two micro-bridge ring devices.

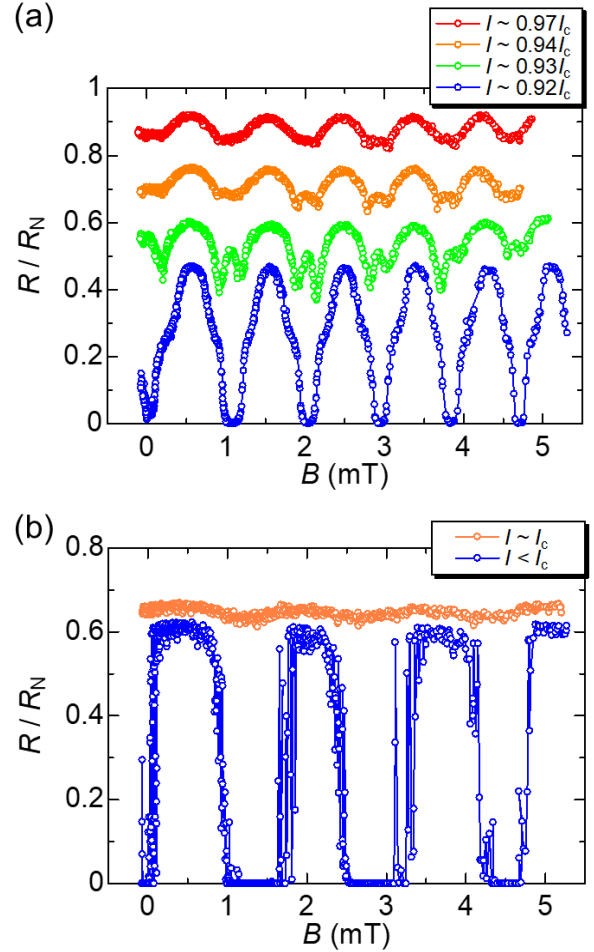

**Fig. 3.** Magnetic field dependence of resistance at 2.5 K obtained with **(a)** micro-bridge ring device and **(b)** simple ring device. The curves in (a) are shifted to the vertical direction for clarity. The vertical axes are normalized by the simple state resistance $R_N$.

Similar resistance oscillations in dc-SQUIDs are explained by the resistively and capacitively shunted junction (RCSJ) model[3]. Based on the theoretical model in Refs. 22)-24), the phase of the resistance oscillation changes from 0 to $\phi_0/2$ as the bias current is increased. This is caused by the LC resonance in the SQUID loop (L: inductance of the loop), which can be excited by a large bias current[3]. In the present experiments, however, such a crossover of the phase was not observed. Thus, the complex resistance oscillations observed in the micro-bridge ring cannot be fully explained by the above model.

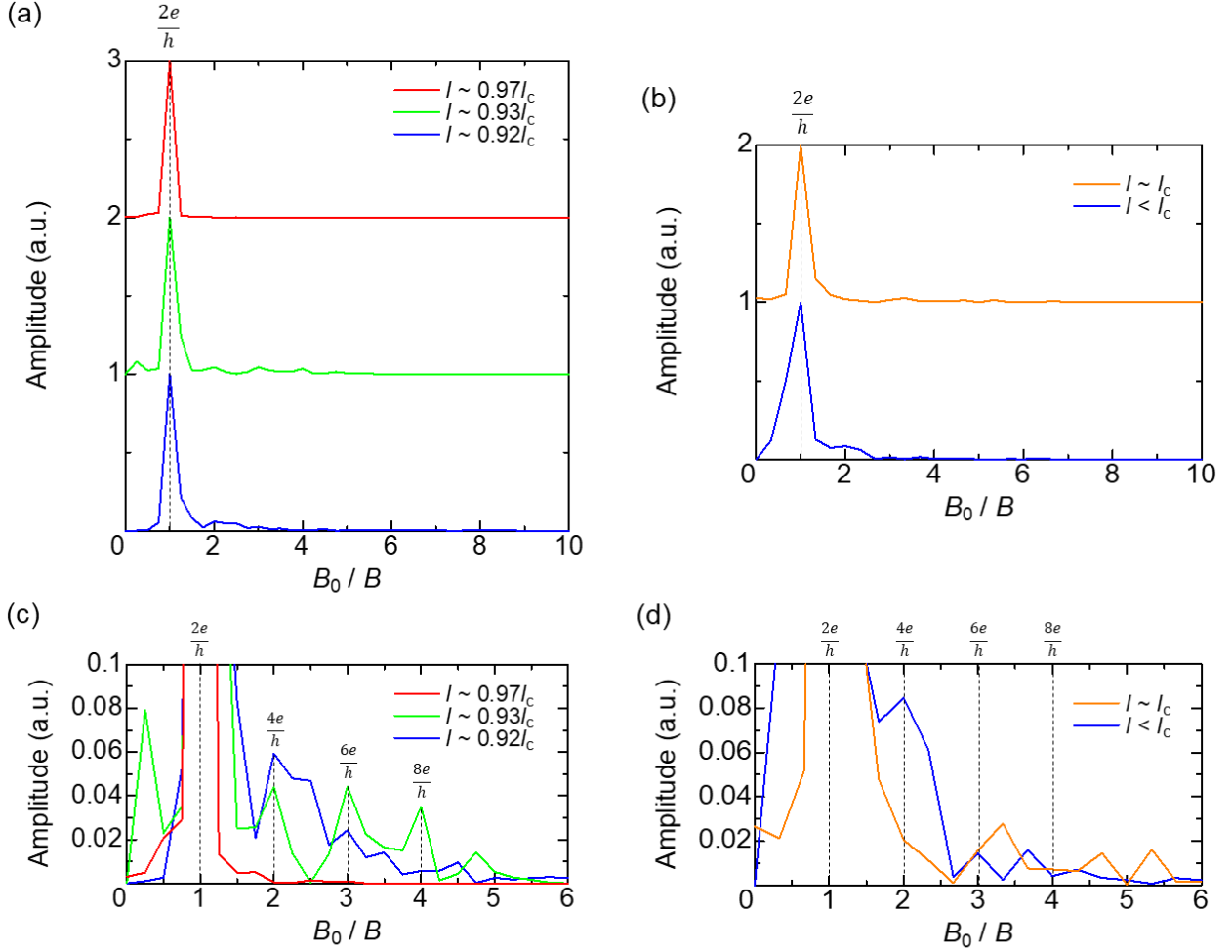

**Fig. 4:** FFT spectra of the resistance oscillations obtained from **(a)** micro-bridge ring and **(b)** simple ring device. The vertical axis is normalized by the peak height at $1/B_0$. The horizontal axis is also normalized by the frequency corresponded to $1/B_0$. The curves are shifted to the vertical direction for clarity. **(c), (d)** The closeup views of **(a)** and **(b)** in the small amplitude region.

To further evaluate the higher harmonic oscillations in Fig. 3(a), we performed FFT analysis.

Figures 4(a) and 4(b) show FFT spectra obtained from the resistance oscillations in the micro-bridge ring and simple ring devices, respectively. All the FFT spectra have a sharp peak at a frequency of $1/B_0$, i.e., $(h/2e)^{-1}$. On the other hand, small satellite peaks can be seen in higher $1/B$ values. In order to focus on the satellite peaks, we show the small amplitude region in Figs. 4(c) and 4(d). Only for the micro-bridge ring device, a few higher harmonic terms were clearly obtained; especially, for $I = 0.93I_c$, up to fourth (i.e., $(h/8e)^{-1}$) harmonic oscillations can be seen in the FFT spectrum. Although the higher harmonic components $(h/4e)^{-1}$ and $(h/6e)^{-1}$ were also observed in the simple ring when the bias current was smaller than $I_c$ (see the blue curve in Fig. 4(d)), these originate from a square wave shape of the resistance oscillation between zero and finite values (see the blue curve in Fig. 3(b)). A similar situation is realized even in the micro-bridge device when the bias current is $0.92I_c$: the FFT spectrum has peaks at $(h/4e)^{-1}$ and $(h/6e)^{-1}$ (see the blue curve in Fig. 4(c)). In this paper, we regard the higher harmonic components obtained from resistance oscillations including the zero resistance regime as the artificial one. The clear difference between the simple ring and micro-bridge ring devices can be seen in the current regime where the resistance does not reach zero, namely the light green curves in Figs. 3(a), 4(a), and 4(c). The resistance maxima appear not only at $(n+1/2)\phi_0$ but also at $n\phi_0$, where $n$ is an integer. Such a complex oscillation pattern results in higher harmonics up to the fourth component and is qualitatively different from the one originating from the artificial reason mentioned above.

Since the superconductivity in the simple ring device is suddenly broken at $I_c$ (see Fig. 2), it is much more difficult to observe clear resistance oscillations than in the micro-bridge ring device. We tried to see resistance oscillations with higher harmonic terms even in the simple ring device by tuning the measurement current and also by using several different ring devices, but such higher harmonic oscillations have never been observed. In addition, to the best of our knowledge, higher harmonic components have not been reported in previous superconducting Nb ring devices[25-30].

Harmonic oscillations with a frequency of $(h/4e)^{-1}$ were reported in Al-based ring-shaped devices[8,31,32] where the lateral size is comparable to $\xi$ of Al. According to Refs. 31) and 32), the second harmonic oscillations can be induced by a phase coupling between the two superconductors via quasiparticles and also by a confinement of quasiparticles to the ring through Andreev reflections. However, the above two effects become much less essential in the present device where the lateral size is much greater than the coherence length $\xi$ of Nb (~38 nm[33]). Since the distance $d$ between the two weak links is about 80 times longer than $\xi$ of Nb, the interaction between the two junctions mediated by Cooper pair[7] or quasiparticle[6,31,32] is not expected. At the moment, there is no theoretical expression on higher harmonic oscillations in a superconducting ring device where $\xi \ll d$. Further theoretical researches are required to elucidate the detailed picture of higher harmonic behavior, especially for the case of $\xi \ll d$.

In conclusion, we measured the magnetic field dependence of resistance in two types of ring-shaped superconducting Nb devices. In a simple Nb ring device, a resistance oscillation with a period of $h/2e$ was observed. In a micro-bridge Nb ring device, on the other hand, higher harmonic resistance oscillations were detected only when the measurement current was well-tuned at around $I_c$. Based on the FFT analysis, the higher harmonics can be seen up to the fourth harmonic component, i.e., $(h/8e)^{-1}$. The second harmonic component $(h/4e)^{-1}$ has been reported so far in Al-based rings where the device size is comparable to $\xi$ of the superconducting Al. On the other hand, the present work has demonstrated that the higher harmonic resistance oscillations can be realized even when the distance between the two SNS junctions is much longer than $\xi$. This result would pave the way for deep understanding of the superconducting interference effect especially in devices with short $\xi$, and would also be useful for a magnetic field sensor as mentioned in Ref. 8).

**Acknowledgements**

We thank S. Iwakiri and T. Ohta for fruitful discussion. This work was supported by JSPS KAKENHI (Grant Numbers JP17K18756, JP20H02557, JP20J20229), JST FOREST (Grant Number JPMJFR2134), the Mazda Foundation, Shimadzu Science Foundation, Yazaki Memorial Foundation for Science and Technology, SCAT Foundation, and the Murata Science Foundation, and the Asahi glass foundation.